\documentclass[%
 reprint,
superscriptaddress,
 amsmath,amssymb,
 aps,
]{revtex4-2}

\usepackage{graphicx}% Include figure files
\usepackage{dcolumn}% Align table columns on decimal point
\usepackage{bm}% bold math
\usepackage{caption}
\begin{document}

\title{PFAS-induced modulation of vesicle diffusiophoresis}% Force line breaks with \\

\author{Seungsu Han}%
\affiliation{Department of Mechanical and Aerospace Engineering, University at Buffalo, The State University of New York, Buffalo, New York 14260, USA}

\author{Gbassey S. A. Oteme}%
\affiliation{Department of Materials Design and Innovation, University at Buffalo, The State University of New York, Buffalo, NY 14260, USA}

\author{Prathima Nalam}%
\affiliation{Department of Mechanical and Aerospace Engineering, University at Buffalo, The State University of New York, Buffalo, New York 14260, USA}
\affiliation{Department of Materials Design and Innovation, University at Buffalo, The State University of New York, Buffalo, NY 14260, USA}

\author{Sangwoo Shin}%
\email{\textbf{E-mail:} sangwoos@buffalo.edu}
\affiliation{Department of Mechanical and Aerospace Engineering, University at Buffalo, The State University of New York, Buffalo, New York 14260, USA}

\begin{abstract}
Understanding how per- and polyfluoroalkyl substances (PFAS) partition into biological membranes and alter their properties is central for elucidating PFAS toxicity. Here, we experimentally investigate how the partitioning of anionic PFAS affects the surface charge of phospholipid vesicles and consequently, their electrokinetically driven vesicle motion, in particular diffusiophoresis. We find that PFAS partitioning enhances vesicle diffusiophoresis, consistent with a model combining the Nernst-Planck and Langmuir adsorption equations. Desorption experiments, together with quartz crystal microbalance with dissipation monitoring (QCM-D), further reveal that substantial lipid loss occurs during PFAS partitioning. We attribute this loss to lipophobic interactions between the fluorocarbon chains of PFAS and the alkyl chains of lipids. Our findings provide new mechanistic insights into how PFAS may disrupt biological systems.

\end{abstract}

\maketitle

%\tableofcontents

\section{Introduction}
Per- and polyfluoroalkyl substances (PFAS), commonly referred to as “forever chemicals,” are a group of surfactants characterized by their fluorocarbon chains. The high electronegativity of fluorine gives these compounds exceptional chemical stability, making them highly resistant to decomposition \cite{samarasinghe2025pfas}. Consequently, the widespread occurrence and persistence of PFAS have raised substantial environmental and public health concerns \cite{panieri2022pfas}. 

In this regard, considerable effort has been put forth to understand the impacts of PFAS on biological systems. Interactions between PFAS and phospholipid membranes have gained particular attention due to their shared amphiphilic characters and the biological importance of membranes \cite{Shen2020,zheng2025molecular,Naumann2022,Nouhi2018,Wang2018,Jbeily2018,Fitzgerald2018,Xie2010}. Recent studies, for instance, have showned that PFAS readily partition into lipid membranes \cite{Shen2020,zheng2025molecular}, thereby altering membrane properties \cite{Nouhi2018,Wang2018,Naumann2022,Jbeily2018}. These findings highlight the potential for PFAS exposure to disrupt biological processes and the need to better understand PFAS-membrane interactions. 

In this study, we investigate how PFAS partitioning alters the electrostatic properties of phospholipid vesicles and, consequently, their electrokinetic motion, namely diffusiophoresis. Diffusiophoresis is the transport of colloidal particles driven by electrolyte concentration gradients, and has been implicated in diverse biological processes, including biomolecular phase separation \cite{doan2024diffusiophoresis,hafner2024reaction,buell2025induction}, drug delivery \cite{shin2019osmotic,doan2021confinement,lee2021diffusiophoresis,ganguly2023diffusiophoresis}, liposome transport \cite{rasmussen2020size,chakra2023continuous,sear2019diffusiophoresis}, motor protein transport \cite{ramm2021diffusiophoretic}, biological pattern formation \cite{alessio2023diffusiophoresis,mirfendereski2026imperfect}, spindle formation \cite{shim2024motorless}, and bacterial motion \cite{Doan_2026, doan2020trace}. This mode of transport is particularly relevant in living systems, which operate far from equilibrium and commonly sustain electrolyte concentration gradients \cite{yamahira2020stepwise}. 

\begin{figure}[b!]
	\centering
	\includegraphics[width=8cm]{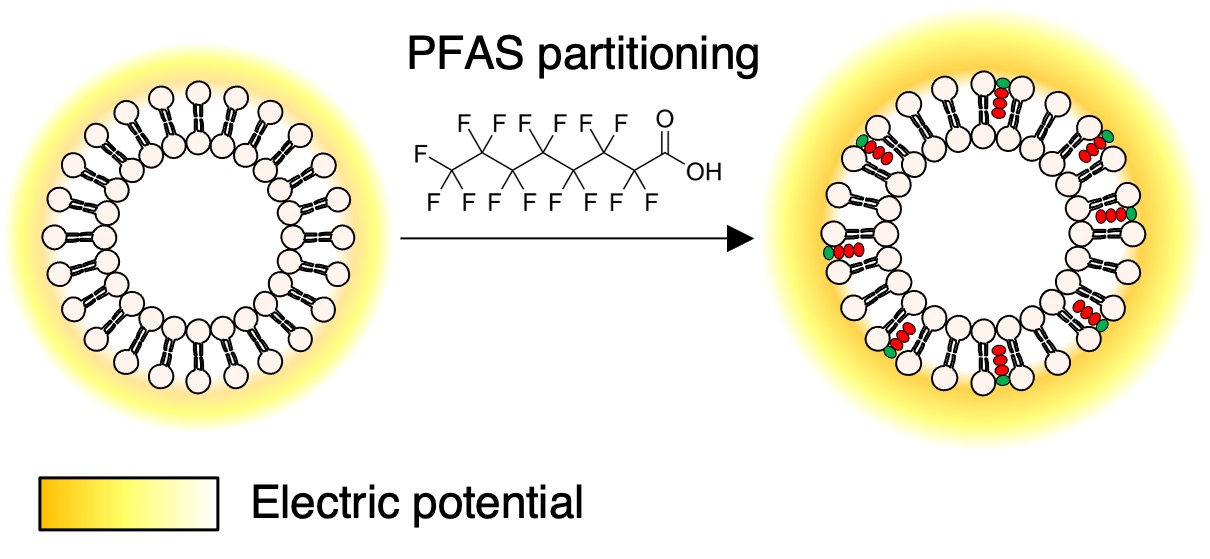}
	\caption{
		\textbf{PFAS partitioning increases zeta potential of phospholipid vesicles.} Hydrophobic fluorocarbon tail of a PFAS molecule induces spontaneous insertion into phospholipid membrane to minimize contact with water while its anionic headgroup contributes to more negative vesicle zeta potential.
	           }
	\label{fig:schematics}
\end{figure}

Additionally, directional vesicle trafficking plays a key role in essential cellular processes \cite{cui2022vesicle}, such as exocytosis \cite{greenwalt2006and} and neurotransmission \cite{eiden2000vesicular}. Despite its biological importance, consequences on the vesicle transport induced by ionic PFAS partitioning remain unexplored. Here, we examine how PFAS partitioning affects diffusiophoretic vesicle transport by dynamically coupling PFAS-induced surface charge modulation and vesicle diffusiophoresis under concurrent salt and PFAS gradients. Given that most PFAS species, including perfluorohexanoic acid (PFHxA), perfluorooctanoic acid (PFOA), perfluorononanoic acid (PFNA), perfluorodecanoic acid (PFDA), and perfluorooctanesulfonic acid (PFOS), are deprotonated at physiological pH \cite{wu2020exploring}, we hypothesize that their association with plasma membranes alters the electrokinetic characteristics of the membranes, as reflected in their zeta potential (\textbf{Fig. \ref{fig:schematics}(a)}). As a result, PFAS-partitioned vesicles are expected to undergo stronger diffusiophoretic migration than PFAS-free vesicles under identical solute gradients. 

\section{Results and discussion}
\subsection{PFAS partitioning impacts zeta potential of lipid membranes}
\begin{figure}
	\centering
	\includegraphics[width=7.5cm]{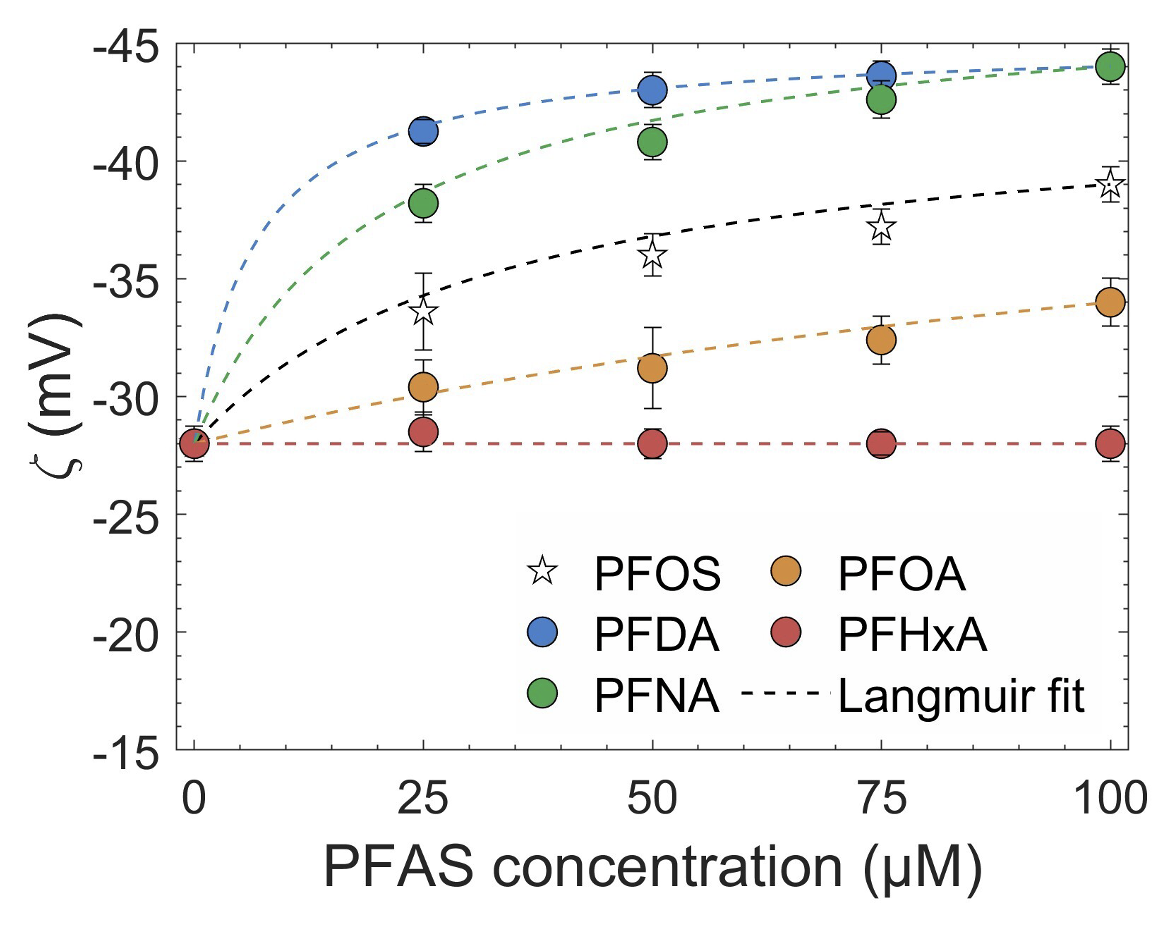}
	\caption{
		\textbf{Vesicle zeta potential becomes more negative with increasing PFAS concentration.} PFAS species with longer fluorocarbon chains induce larger shifts in the vesicle zeta potential. Dashed lines indicate fits of the Langmuir adsorption model to the measured PFAS concentration-dependent zeta potentials.
	           }
	\label{fig:zeta}
\end{figure}

The impact of external PFAS concentration on the vesicle zeta potential is shown in \textbf{Fig. \ref{fig:zeta}}. In the absence of PFAS, model plasma membrane vesicles composed of DOPC and Rh-DPPE (99.5:0.5) exhibit a baseline zeta potential of –28 $\pm$ 0.8 mV in 1 mM NaCl solution. The size distribution of the measured GUVs is shown in Fig. S1. Upon the addition of PFAS at \textmu M concentrations, the zeta potential becomes progressively more negative as the PFAS concentration increases. 

The observed zeta potential change is attributed to the partitioning of anionic PFAS into the vesicle membranes rather than to charge screening \cite{kirby2004zeta} or size effect caused by fission \cite{HAN202513908}. Charge screening contributions are limited by the low PFAS concentrations, while size-dependent effects are negligible in the thin Debye layer limit ($a/\lambda_D\gg 1$), where $\lambda_D$ is the Debye length and $a$ is the vesicle radius. To quantify PFAS-membrane interactions, we fit the measured zeta potentials as a function of PFAS concentration using a Langmuir adsorption model. Assuming that PFAS partitioning is insensitive to the modest variation in the ionic strength over the range 0.1–10 mM \cite{aka2026remediation}, we express the vesicle zeta potential, $\zeta$, as 
\begin{equation}
\zeta = a_{0} + a_{1} \log{\lambda_{D}} + (a_{2}-g) \dfrac{K c_{P}}  {1+K c_{P}} ,
\label{eqn:model}
\end{equation}
where $K$ is the Langmuir absorption constant and $c_{P}$ is the PFAS concentration. The first two terms describe classical Debye screening \cite{kirby2004zeta}, while the last term captures the competing effects of anionic PFAS partitioning and PFAS-induced lipid loss. The parameters $a_{0}$ and $a_{1}$ characterize ionic screening and are determined independently from zeta potential measurements of DOPC/Rh-DPPE vesicles over a range of NaCl concentrations. The parameters $a_{2}$ and $K$ are obtained by fitting the PFAS-dependent zeta potential values to the Langmuir adsorption model. The model parameter $g$ accounts for the loss of anionic lipids promoted by PFAS partitioning and is determined from the measured change in zeta potential after vesicle rinsing, which will be discussed later in Section C. Overall, our model agrees well with the experimental data, as indicated by the dashed lines in \textbf{Fig. \ref{fig:zeta}}.

The PFAS-induced shift in the zeta potential depends on both fluorocarbon chain length and headgroup chemistry \cite{ebert2020membrane}. In general, longer-chain PFAS have a greater impact on zeta potential because they partition more strongly into lipid membranes \cite{chen2025deriving}. This tendency is attributed to their extended hydrophobic fluorocarbon surface area, which favors partitioning away from water and into the membrane \cite{Shen2020,zheng2025molecular}. At $c_{p}=100$ \textmu M, PFDA, which is the longest-chain PFAS examined here (C10), shifts the vesicle zeta potential from –28 mV to –44 mV, whereas PFOA (C8) shifts the zeta potential only to –34 mV. PFNA (C9), partitions weaker than PFDA \cite{ebert2020membrane}, producing a smaller shift. However, at $c_{p}=100$ \textmu M, PFNA and PFDA yield comparable zeta potentials, suggesting saturation of the membrane partitioning capacity where increasing the PFAS concentration does not contribute to accumulating more PFAS in the membrane. At the lower extreme, PFHxA, which is the shortest-chain PFAS examined here (C6), produces no measurable change in vesicle zeta potential over the tested concentration range. Its relatively weak affinity for lipid membranes likely prevents measurable accumulation within the lipid bilayer at \textmu M concentrations.

Headgroup chemistry also modulates PFAS partitioning. Although PFOS and PFOA share the same fluorocarbon chains of the same length, the sulfonate headgroup of PFOS promotes stronger partitioning than the carboxylate headgroup of PFOA. In fact, the reported water/membrane partition coefficient is approximately one order of magnitude larger for PFOS than PFOA \cite{Xie2010,xie2007effect}. 

\begin{figure*}
	\centering
	\includegraphics[width=\textwidth]{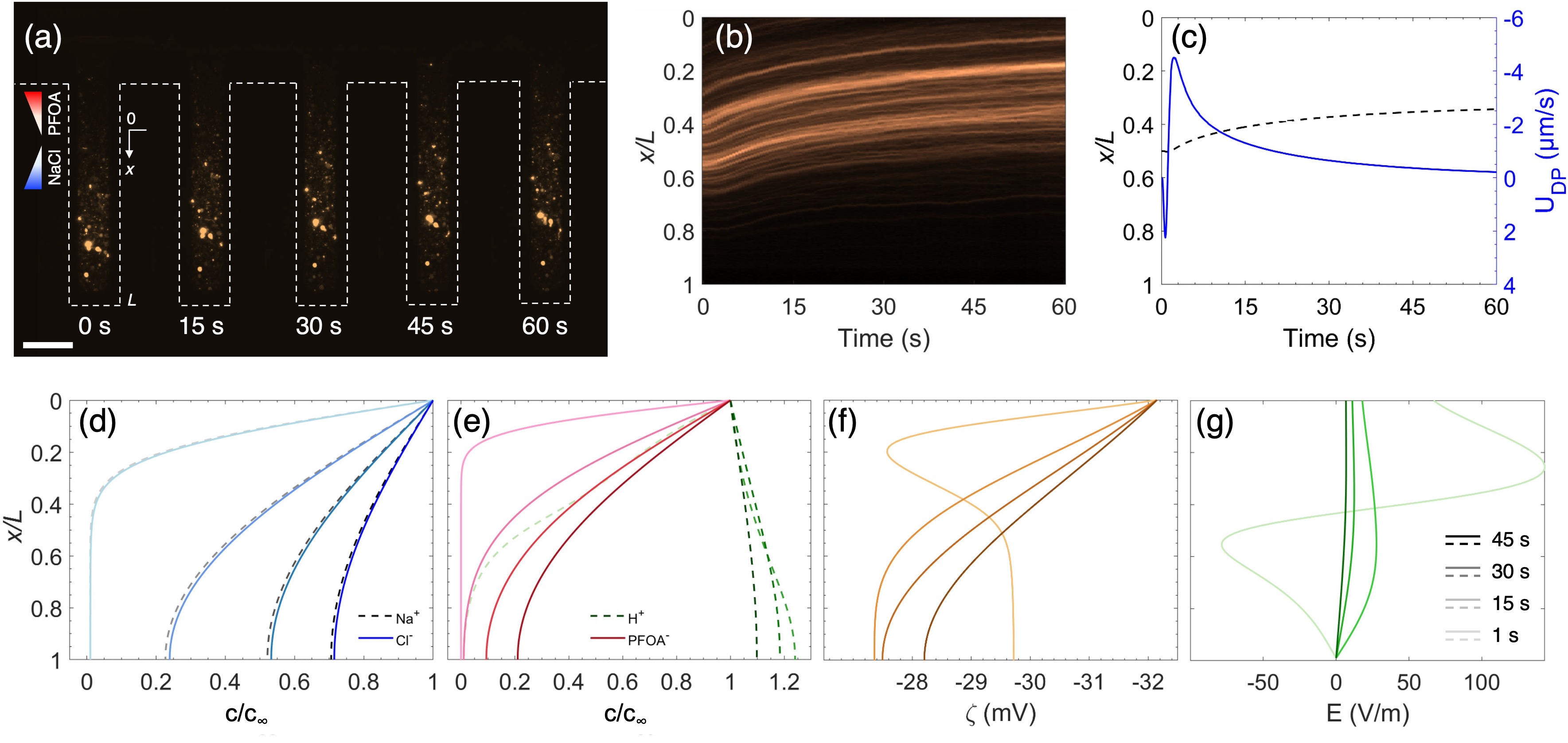}
	\caption{
		\textbf{Diffusiophoresis of giant unilamellar vesicles (GUVs) in dead-end pores under concurrent NaCl and PFOA gradients.} (a) GUVs (orange) are confined within microfluidic dead-end pores (white dashed lines). The pores initially contain 0.1 mM NaCl. At $t=0$ s, a solution containing 10 mM NaCl and 0.1 mM PFOA is introduced at the pore entrance ($x=0$). Scale bar is 50 \textmu m. (b) Kymograph of vesicle trajectories along the pore direction, where $L$ denotes the pore length. (c) Calculated diffusiophoretic velocity (solid blue line) and vesicle trajectory (dashed black line) obtained from the spatiotemporal ion concentrations (d, e). Local vesicle zeta potential (f), and the electric field induced by the diffusion of ionic species (g). The local zeta potential depends on the PFAS concentration and ionic strength. PFOA is assumed to be fully deprotonated at the pore entrance such that the proton and PFOA anion concentrations are equal to the imposed PFOA concentration. Curve shades from light to dark correspond to $t=1$, 15, 30, and 45 s, respectively. 
	           }
	\label{fig:exp}
\end{figure*}
%\FloatBarrier

\subsection{PFAS-partitioning dynamically regulates diffusiophoresis}
Having established that PFAS partitioning directly alters the vesicle zeta potential, we next examine its effect on diffusiophoretic transport under dynamic conditions that resemble non-equilibrium biological environments by performing experiments under concurrent gradients of NaCl and PFAS concentrations. The NaCl gradients drive diffusiophoretic transport, whereas the diffusing PFAS continuously partition into the lipid membrane and dynamically altering the vesicle zeta potential. This coupled gradient configuration allows us to test the hypothesis that transient PFAS partitioning modulates vesicle diffusiophoresis.

Diffusiophoresis experiments were conducted in a microfluidic dead-end pores (\textbf{Fig. \ref{fig:exp}(a, b)}) \cite{ault2017diffusiophoresis, shin2017low, duong2026salt}. The pores outlined by white dashed lines initially contained GUVs (orange) dispersed in 0.1 mM NaCl. At $t=0$~s, a solution containing 10 mM NaCl and 100 \textmu M PFAS anions, together with the corresponding protons released upon PFAS deprotonation, was introduced at the pore entrance. This establishes a step change in electrolyte concentration at the pore entrance ($x=0$) that propagated into the pore. As the solute molecules diffuse, the electrolyte concentration evolves toward error function form and gradually relaxes to equilibrium. The resulting NaCl gradients drive the negatively charged GUVs toward the pore entrance by diffusiophoresis. This directional migration of GUVs in dead-end pores is captured in \textbf{Fig. \ref{fig:exp}(a)}, where their trajectories are visualized as a kymograph in \textbf{Fig. \ref{fig:exp}(b)}.

We predict the vesicle trajectory by coupling the Nernst-Planck equation with the zeta potential relation in Eq. 1 (\textbf{Fig. \ref{fig:exp}(c)}). The spatiotemporal evolution of the ionic concentrations (\textbf{Fig. \ref{fig:exp}(d, e)}) is obtained by numerically solving the one-dimensional species conservation equation
\begin{equation}
\frac{\partial{c_{i}}}{\partial{t}} + \nabla \cdot \textbf{j}_{i}=0,
\label{eqn:nernst1}
\end{equation}
where $c_{i}$ and $\textbf{j}_{i}$ denote the concentration and flux of ionic species $\it{i}$, respectively. The ionic flux is given by 
\begin{equation}
\textbf{j}_{i}=-D_{i}\nabla c_{i} + \frac{D_{i}z_{i}e}{k_{B} T}c_{i}\textbf{E},
\label{eqn:nernst2}
\end{equation}
where $D_i$ and $z_i$ are the diffusion coefficient and valence of species $i$, respectively, $e$ is the elementary charge, $k_{B}$ is the Boltzmann constant, $T$ is the temperature, and \textbf{E} is the electric field. Diffusion of ionic species generates a diffusion electric field
\begin{equation}
\textbf{E}=\frac{k_{B} T}{e} \frac{\Sigma_{i}z_{i}D_{i}\nabla c_{i}}{\Sigma_{i}z_{i}^2D_{i} c_{i}}.
\end{equation}
Finally, diffusiophoretic velocity $U_{DP}$ for multi-ion gradients is \cite{chiang2014multi,gupta2019diffusiophoretic}
\begin{equation}
\mathbf{U}_{DP} =  \frac{\epsilon}{8\mu}\frac{\Sigma_{i}z_{i}^2 \nabla c_{i}}{\Sigma_{i}z_{i}^2 c_{i}}\zeta^2+\frac{\epsilon k_{B} T}{e\mu}\frac{\Sigma_{i}z_{i}D_{i}\nabla c_{i}}{\Sigma_{i}z_{i}^2D_{i}c_{i}}\zeta,
\end{equation}
where $\mu$ is the fluid viscosity and $\epsilon$ is the electric permittivity. Subsituting the calculated ionic concentration fields and vesicle zeta potential into the above expression yields the diffusiophoretic velocity $\mathbf{U}_{DP}$. The corresponding vesicle trajectory is obtained by solving $dx_p/dt=\mathbf{U}_{DP}(x_p(t),t)$, where both the velocity and trajectory are presented in \textbf{Fig. \ref{fig:exp}(c)}. 

The vesicles initially exhibit a zeta potential of approximately –29.8~mV in 0.1~mM NaCl. Introducing concentrated NaCl rapidly screens the vesicle surface charge and reduces the magnitude of the zeta potential (\textbf{Fig. \ref{fig:exp}(f)}). As PFOA diffuses into the pore and partitions into the lipid membranes, its anionic charge progressively offsets this screening and shifts the zeta potential to more negative values. By $t=15$~s, PFOA partitioning measurably increases the zeta potential in regions where the local PFOA concentration is sufficiently high (\textbf{Fig. \ref{fig:exp}(f)}).

\begin{figure}
	\centering
  \includegraphics[width=8.5cm]{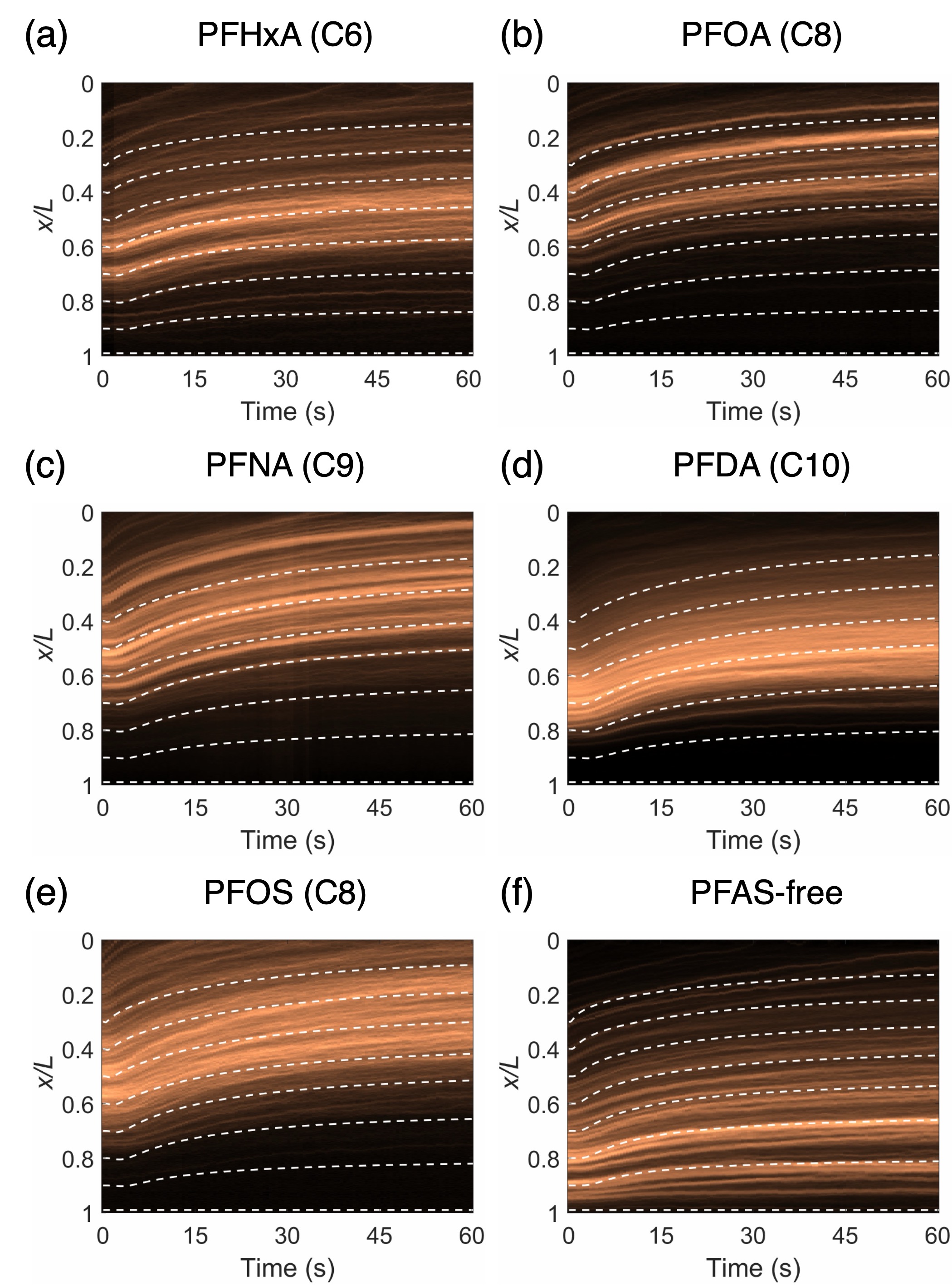}
    \caption{\textbf{Experimental and predicted vesicle trajectories under concurrent NaCl and PFAS concentrations.} Experimental kymographs showing vesicle trajectories (orange) along the pore in the $x$ direction. The corresponding model predictions are shown as white dashed lines. Each line represents movement of vesicles over the path of the pore. Kymographs show vesicle trajectories for different PFAS species of (a) PFHxA; (b) PFOA; (c) PFNA; (d) PFDA; (e) PFOS; (f) PFAS-free control case with NaCl gradient only. 
}
    \label{fig:kymo}
\end{figure}

The experimental vesicle trajectories are compared with our model prediction in \textbf{Fig. \ref{fig:kymo}}. The orange lines represent experimentally measured vesicle trajectories within the pore, showing their migration toward the pore entrance ($x=0$), whereas the white dashed lines are non-Brownian trajectories calculated from our transport model. Vesicle trajectories vary among PFAS species because their membrane partitioning and diffusion coefficients differ. Each experiment was performed for 60 seconds after the introduction of the salt solutions. Longer-chain PFAS exhibit stronger membrane partitioning, producing larger shifts in vesicle zeta potential and steeper diffusiophoretic trajectories. Accordingly, all tested PFAS except PFHxA enhance vesicle diffusiophoresis relative to the PFAS-free control, in which transport is driven solely by the NaCl gradient (\textbf{Fig. \ref{fig:kymo}(f)}). The PFHxA trajectory is indistinguishable from the control trajectory (\textbf{Fig. \ref{fig:kymo}(a})) because its short fluorocarbon chain results in negligible membrane partitioning.

Direct comparison of the experimental vesicle trajectories and the model predictions demonstrate that the coupled transport-partitioning model accurately captures vesicle diffusiophoresis under dynamic PFAS exposure. In particular, the agreement between the experimental and predicted kymographs indicates that the Langmuir adsorption model adequately describes time-dependent PFAS partitioning, consistent with a single-site partitioning mechanism. This agreement supports applying the model to conditions in which PFAS concentrations vary over time and suggests its utility for predicting vesicle transport in transient biochemical environments \cite{yamahira2020stepwise}.

\subsection{Lipid loss induced by PFAS partitioning}

\begin{figure}
	\centering
	\includegraphics[width=7.2cm]{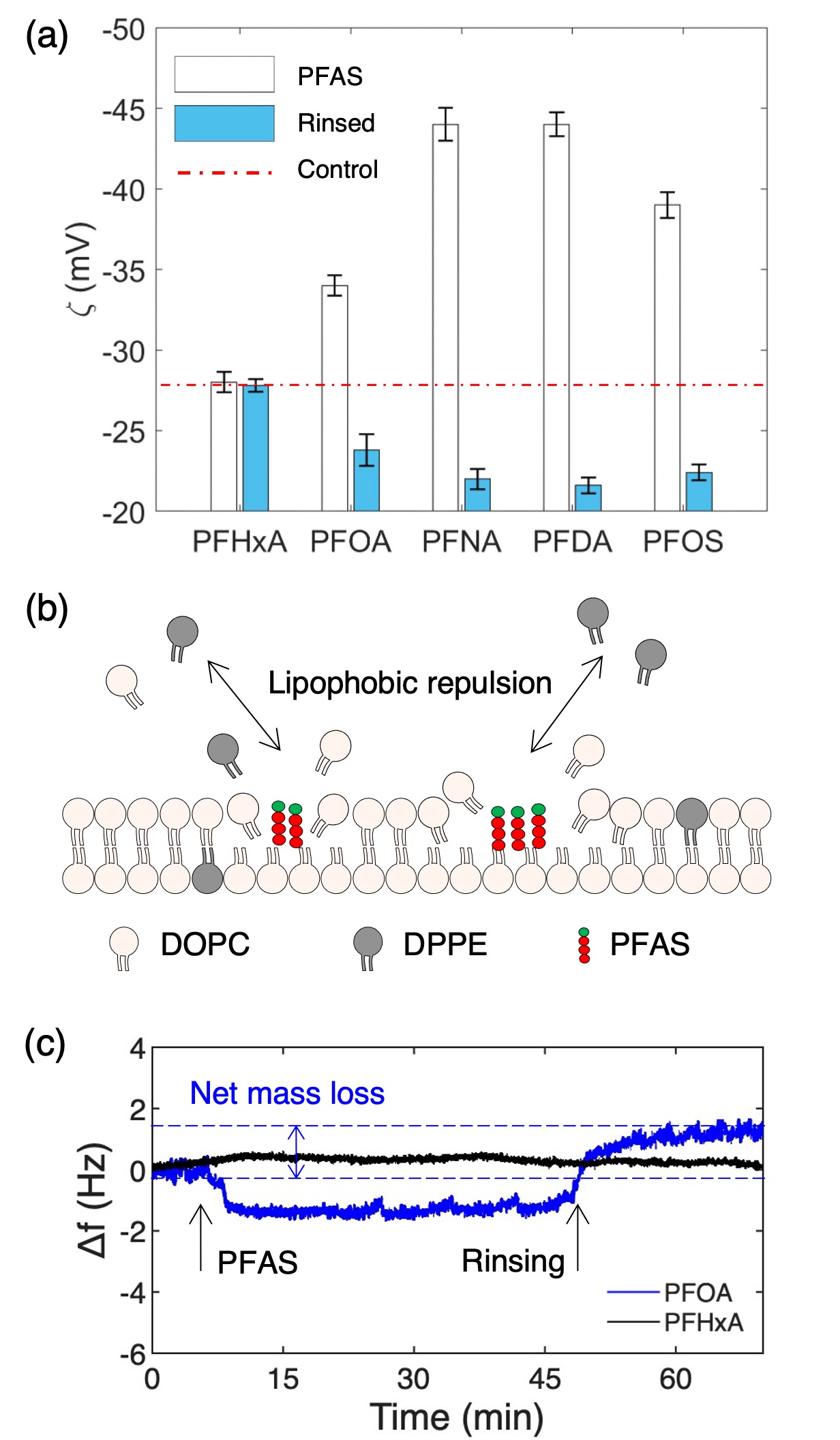}
	\caption{
		\textbf{PFAS partitioning induces lipid loss and consequently reduces vesicle zeta potential after rinsing.} (a) Vesicle zeta potential after PFAS exposure (white) and subsequent rinsing (blue). (b) Proposed mechanism of PFAS-induced lipid loss arising from unfavorable lipophobic interaction between PFAS fluorocarbon chains and lipid alkyl chains. DPPE is removed more rapidly because its saturated alkyl tails interact less favorably with the PFAS fluorocarbon chains, and the resulting difference in lipid-removal rates alters the membrane composition after rinsing. Vesicle zeta potential after PFAS exposure (white) and subsequent rinsing (blue). (c) Change in the resonant frequency of a silica disk coated with a supported DOPC lipid bilayer during PFAS partitioning and desorption. At $t=10$~min, 0.1~mM PFOA (blue) or (black) PFHxA in 10~mM NaCl is introduced. At $t=45$~min, PFAS-free 10~mM NaCl is introduced to rinse the membrane. 
	    }
	\label{fig:loss}
\end{figure}

To evaluate the reversibility of PFAS partitioning and the potential for hysteresis in vesicle electrokinetic behavior, we investigated PFAS desorption from lipid membranes. Vesicles were first exposed to PFAS at $c_{P}=100$ \textmu M and then rinsed thoroughly with 0.1 mM NaCl for at least 1 h before subsequent zeta potential measurements. After rinsing, the magnitude of the vesicle zeta potential unexpectedly fell below its initial value of --28 mV (\textbf{Fig. \ref{fig:loss}(a)}). Moreover, the magnitude of this post-rinse reduction increased with PFAS chain length, indicating that PFAS species with stronger membrane partitioning produce larger persistent changes in vesicle surface charge.

We attribute the post-rinse decrease in zeta potential to lipid loss driven by lipophobic interactions between PFAS fluorocarbon chains and lipid alkyl chains. Fluorination gives PFAS both hydrophobic and lipophobic character, favoring configurations that minimize interfacial free energy \cite{Nouhi2018,krafft2001fluorocarbons}. These unfavorable PFAS-lipid interactions promote lipid removal from vesicle membranes. Because lipid loss is coupled to PFAS partitioning, its extent increases with partitioning strength \cite{Nouhi2018}. This relationship explains why longer-chain PFAS, such as PFNA and PFDA, produce larger post-rinse reductions in zeta potential (\textbf{Fig. \ref{fig:loss}(b)}).

Noticeably, the observed reduction in zeta potential further suggests selective lipid loss. Specifically, the anionic Rh-DPPE lipids that confer surface charge must be lost to a greater extent than the zwitterionic DOPC lipids, thereby lowering the DPPE-to-DOPC ratio relative to its initial value. We attribute this preferential loss to unfavorable interactions between the saturated alkyl chains of DPPE and the fluorocarbon chains of PFAS, rather than to differences in lipid headgroup chemistry. Consistent with this interpretation, our previous free energy calculations from molecular dynamics simulations showed that PFAS interact more unfavorably with saturated lipid tails than with unsaturated tails \cite{HAN202513908}.

Quartz crystal microbalance with dissipation monitoring (QCM-D) provided independent evidence of PFAS-induced lipid loss. A DOPC bilayer was initially deposited on the silica disk oscillating at its resonant frequency (Fig. S2). After bilayer formation, the frequency was calibrated to the baseline of 0~Hz (\textbf{Fig. \ref{fig:loss}(c)}), and subsequent changes were used to monitor changes in the mass associated with the lipid membrane. At $t=10$ min, 100 \textmu M PFOA in 10~mM NaCl was introduced to the lipid bilayer. PFOA partitioning into the membrane increased the coupled mass and consequently decreased the sensor’s resonant frequency. At approximately $t=45$~min, the disk was rinsed with PFAS-free 10 mM NaCl. The resulting chemical potential gradient across the membrane and bulk fluid drove residual PFAS out of the lipid bilayer. 

During rinsing, the resonant frequency increased as PFAS mass was removed from the membrane. The frequency subsequently reached a new equilibrium above the initial 0 Hz baseline, indicating a net loss of membrane mass, and therefore lipid removal following PFAS exposure. Together with the zeta potential measurements (\textbf{Fig. \ref{fig:loss}(a)}), the QCM-D results support our interpretation of PFAS-induced lipid loss. We confirm that the net mass loss was observed for all tested PFAS except PFHxA (Fig. S3). Consistent with its negligible membrane partitioning, PFHxA produced no measurable frequency shift during either exposure or rinsing (\textbf{Fig. \ref{fig:loss}(c)}).

\section{Conclusion}
By coupling PFAS partitioning with diffusiophoretic transport model, our results demonstrate that partitioning of anionic PFAS increases the vesicle zeta potential with concentration dependence consistent with the Langmuir adsorption model. PFAS chain length and headgroup chemistry governed the magnitude of this change primarily through their effects on membrane partitioning strength. The resulting modulation of vesicle surface charge altered diffusiophoretic migration, a transport mechanism potentially relevant to cellular processes \cite{cui2022vesicle,eiden2000vesicular}. These findings may provide a foundation for future investigations into the biological consequences of PFAS partitioning for cellular transport.

Upon rinsing, the vesicle zeta potential decreased to values lower than their initial level, suggesting preferential loss of anionic Rh-DPPE. We attribute this selectivity to lipophobic interactions between PFAS fluorocarbon chains and lipid alkyl chains, which are more unfavorable for saturated lipid tails. While our model vesicles are binary lipid mixtures, biological plasma membranes are considerably more complex, containing diverse lipid species, cholesterol-rich domains, membrane proteins, and receptors \cite{harayama2018understanding}. Nevertheless, our findings provide insight into a potential mechanism by which PFAS may compromise membrane integrity through selective lipid removal. Although the simplified model system does not fully capture the compositional and structural complexity of biological membranes, our results suggest that PFAS-induced lipid redistribution and membrane disruption may also be relevant to biological systems. Further studies using more physiologically representative membrane compositions are needed to establish the extent to which these mechanisms occur in living cells. These findings may warrant future investigations, such as PFAS-induced expulsion in real biological configurations of lipid rafts \cite{simons2004model} or deconstruction of plasma membranes \cite{HAN202513908}. 

Finally, our study suggests a new opportunity for quantifying PFAS-lipid interactions. Because PFAS partitioning prominently alters the electrostatic properties of plasma membranes, the extent of PFAS-lipid association may be inferred from electrokinetic colloidal mobilities. 
\section*{Conflicts of interest}
There are no conflicts to declare.

\acknowledgments{This material is based upon work supported by the National Science Foundation under Grant No. 2237177 and 2238874.}

\appendix

\section{Materials and Methods}

\subsection{Materials}
1,2-dioleoyl-sn-glycero-3-phosphocholine (DOPC) and 16:0 Liss Rhod PE (Rh-DPPE) are purchased from Avanti Research (AL, USA) and dissolved in chloroform (Avantor, PA, USA). Perfluorohexanoic acid (PFHxA), perfluorononanoic acid (PFNA), perfluorodecanoic acid (PFDA), and perfluorooctane sulfonate (PFOS) were purchased from Cayman Chemical (MI, USA). Perfluorooctanoic acid (PFOA), NaCl, and polyvinyl alcohol (PVA) were purchased from Sigma-Aldrich (MO, USA). Polydimethylsiloxane (PDMS; Sylgard 184) was purchased from Dow Corning (MI, USA). 

\subsection{Vesicle preparation}
Giant Unilamellar Vesicles (GUVs) were prepared by bath sonication. 4 mM DOPC lipids were dissolved in chloroform along with 0.5\% Rh-DPPE and deposited in a glass vial. Then, chloroform was evaporated in a desiccator for at least 30 minutes to form a dry lipid film. The film was then hydrated with 10 mM NaCl and sonicated in a bath sonicator (UC 300 Ultrasonic Cleaner, Cole Palmer, IL, USA) for 2 hr until GUVs were formed. The size distribution of GUVs measured by dynamic light scattering (Litesizer 500, Anton Paar, Austria) is presented in \textbf{Fig. S1}. Although the measured distribution includes sub-micron vesicles, the vesicles resolved under our optical microscope and analyzed in the kymograph (\textbf{Fig. 4}) are predominantly larger than 1 \textmu m.

\begin{figure}[b]
	\centering
	\includegraphics[width=7cm]{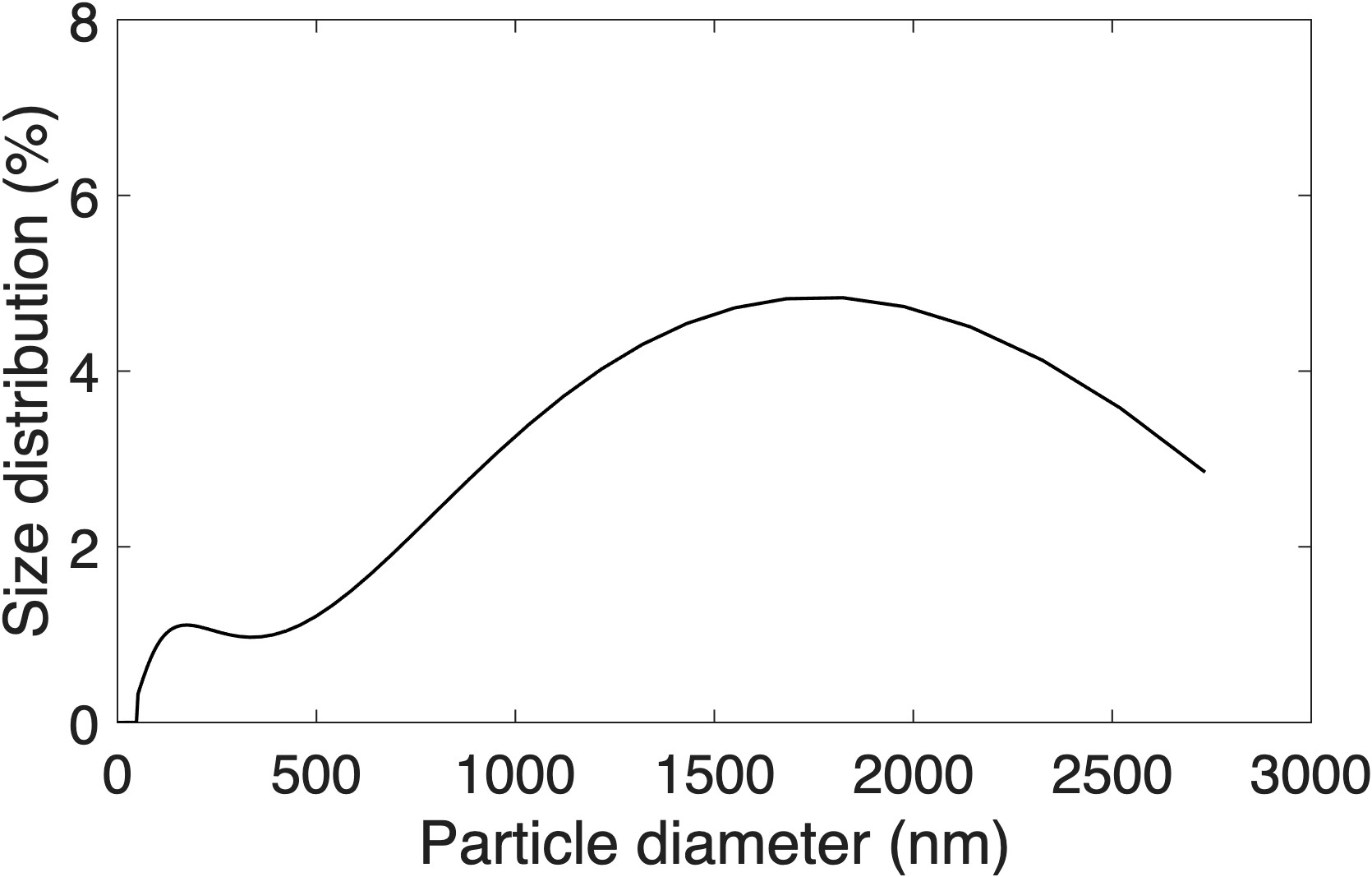}
	\caption*{Fig. S1: 
		\textbf{Size distribution of GUVs measured by dynamic light scattering.}
	           }
	\label{fig:size}
\end{figure}

Small unilamellar vesicles (SUVs) were prepared by probe sonication. A dry lipid film was prepared as described above and hydrated with 100 mM NaCl. The resulting suspension was sonicated using an immersed probe sonicator (Emerson, MO, USA) at 10\% of its maximum power for 5 min. Metal debris shed from the sonication probe was subsequently removed by centrifugation at 10,000 rpm for 10 min.

\subsection{Diffusiophoresis in microfluidic dead-end channels} 
The PDMS microfluidic channels containing dead-end pores \cite{shin2020diffusiophoretic} were coated with PVA to minimize electrostatic interactions between the fluid and the PDMS surfaces. The PDMS channels were first oxidized by plasma treatment, after which a 1\% (w/v) PVA solution was introduced into the channels and incubated for 10 min. Excess PVA solution was removed by pipetting, and the device was incubated overnight in an oven at 75~\textdegree C. Residual PVA was removed by continuously flushing the channels with 0.1 mM NaCl for at least 1 h. 

For the diffusiophoresis experiments, a solution containing 0.1 mM PFAS and 10 mM NaCl was introduced using a syringe pump (Pump 11 Pico Plus Elite, Harvard Apparatus, MA, USA) into the microfluidic channels, whose dead-end pores initially contained GUVs in 0.1 mM NaCl. The resulting electrolyte concentration gradients drove the negatively charged vesicles by diffusiophoresis. Vesicle motion was imaged using an inverted fluorescence microscope (DMi8, Leica, Germany) equipped with an sCMOS camera (K5, Leica, Germany). The images were analyzed using ImageJ \cite{schindelin2012fiji} and custom MATLAB code.

\subsection{Zeta potential measurement by diffusiophoresis} 
To determine the vesicle zeta potential, GUVs initially suspended in 0.1 mM NaCl within the dead-end pores were exposed to a concentration gradient generated by introducing 10 mM NaCl at the pore entrance \cite{shin2017low}. The measured vesicle trajectories were compared with theoretical predictions generated over a range of zeta potential values, and the value producing the best fit was assigned as the vesicle zeta potential. To quantify the effect of PFAS, the procedure was repeated with the same PFAS concentration added to both the 0.1 mM and 10 mM NaCl solutions. This configuration maintained the NaCl concentration gradient while eliminating the PFAS concentration gradient.

\subsection{Zeta potential model parameters}
The vesicle zeta potential was modeled by combining the contributions of ionic screening and PFAS partitioning:
\begin{equation}
\zeta = a_{0} + a_{1} \log{\lambda_{D}} + (a_{2}-g) \dfrac{K c_{P}}  {1+K c_{P}}.
\label{eqn:model}
\end{equation}
The model parameters used for each PFAS, including the Langmuir partitioning constant $K$, are summarized in Table S1. The values of $a_{2}$ and $K$ were obtained by fitting the measured zeta potentials to the Langmuir adsorption model using a built-in MATLAB optimization routine. Parameters could not be determined for PFHxA because its partitioning-induced effect was not measurable at \textmu M concentrations. Among the perfluorocarboxylic acids, $K$ increased with fluorocarbon chain length, consistent with stronger membrane-partitioning affinity. The ionic screening parameters, $a_{0}=-0.056$ V and $a_{1}=-0.00155$ V, were determined from zeta-potential measurements of DOPC/Rh-DPPE vesicles over the experimental range of NaCl concentrations. 
$\newline$
\begin{table}[h!]
\centering
 \begin{tabular}{c c c c c} 
 \hline
  Parameters&  PFOA& PFNA&  PFDA& PFOS\\[1ex]
 \hline
K (L/mmol)&  5,008& 41,835& 135,605& 20,836\\[0.5ex] 
$a_{2}$ (V)&  -0.0267&-0.0264& -0.02345& -0.02267\\[0.5ex] 
g (V)&  0.01067& 0.0072& 0.0064&0.008\\[0.5ex] 
  \hline
 \end{tabular}
 \caption*{Table S1: Partition coefficients and free parameters used to model PFAS-induced zeta potential modulation.}
\end{table}

\section{Quartz crystal microbalance with dissipation monitoring}

Before $t=0$~min, 100~mM NaCl was continuously flowed over the quartz crystal sensor at 50~\textmu L/min. After the resonant frequency stabilized, the frequency shift was defined as 0~Hz. At $t=15$~min, SUVs suspended in 100~mM NaCl were introduced to form a supported lipid bilayer on the silica surface. Vesicle adsorption and fusion caused a rapid decrease in resonant frequency, followed by a transient undershoot and stabilization at approximately --30~Hz, consistent with lipid bilayer formation. The resulting DOPC bilayer had a measured mass density of 379~ng/cm$^{2}$, in agreement with previously reported values \cite{parkkila2018biophysical}.

\begin{figure}
	\centering
	\includegraphics[width=7cm]{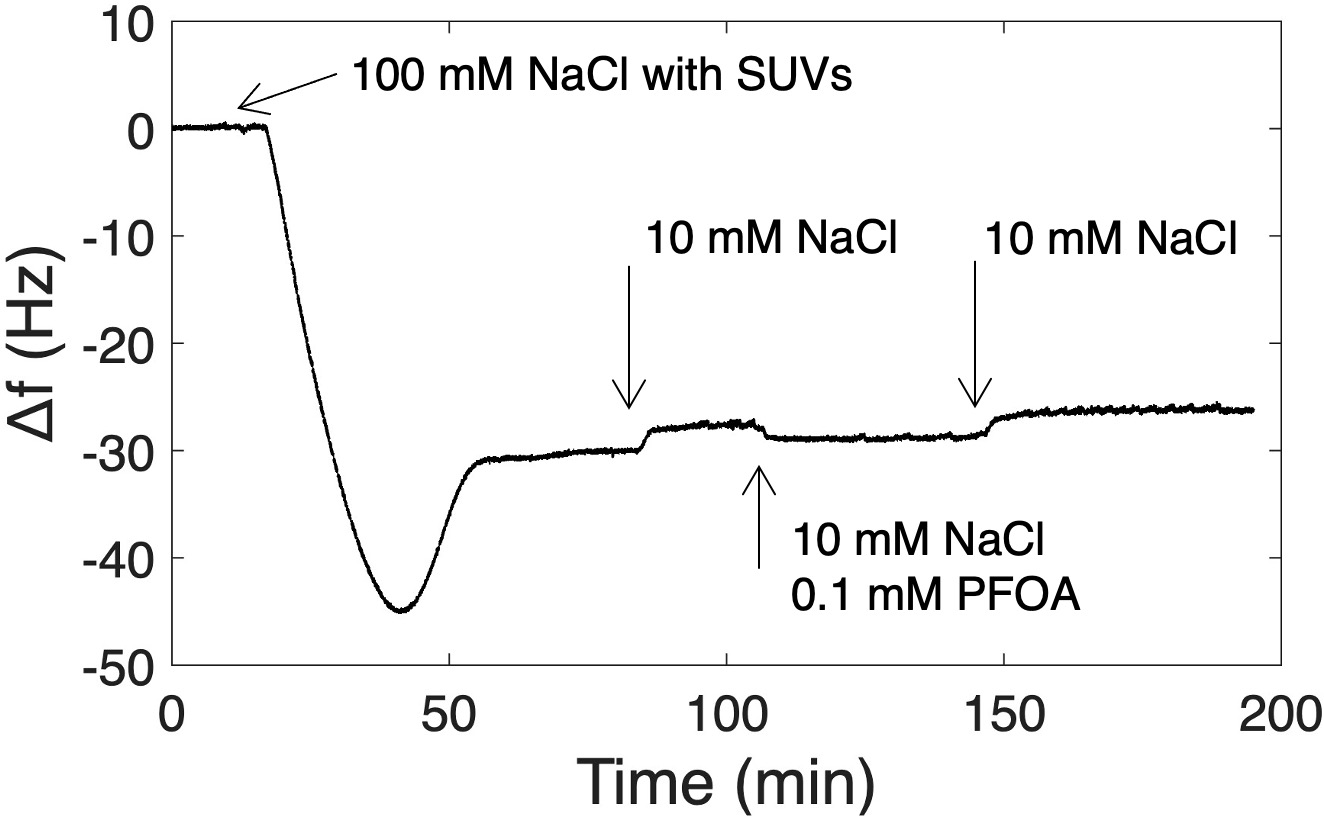}
	\caption*{Fig. S1: 
		\textbf{QCM-D frequency response during lipid bilayer formation, PFAS partitioning, and desorption.
	           }}
	\label{fig:qcm}
\end{figure}
$\newline$

After the supported lipid bilayer stabilized, the flowing solution was switched first to 10~mM NaCl and then to 0.1~mM PFAS in 10~mM NaCl (\textbf{Fig. S2}). The decrease in resonant frequency following PFAS introduction indicated a net increase in the mass coupled to the bilayer, implying that the mass gained through PFAS partitioning exceeded the concurrent mass loss caused by lipid removal. The surface mass density change, $\Delta m$, was calculated from the measured frequency shift, $\Delta f$, using
\begin{equation}
\Delta m = -C \dfrac{\Delta f}{n},
\label{eqn:sauerbrey}
\end{equation}
where $C=17.7$~ng$\cdot$cm$^{-2}$$\cdot$Hz$^{-1}$ is the mass-sensitivity constant and $n$ is the harmonic number. The values of $\Delta m$ measured during partitioning and desorption for the different PFAS are presented in \textbf{Fig. S3}.
$\newline$

\begin{figure} [h]
	\centering
	\includegraphics[width=7cm]{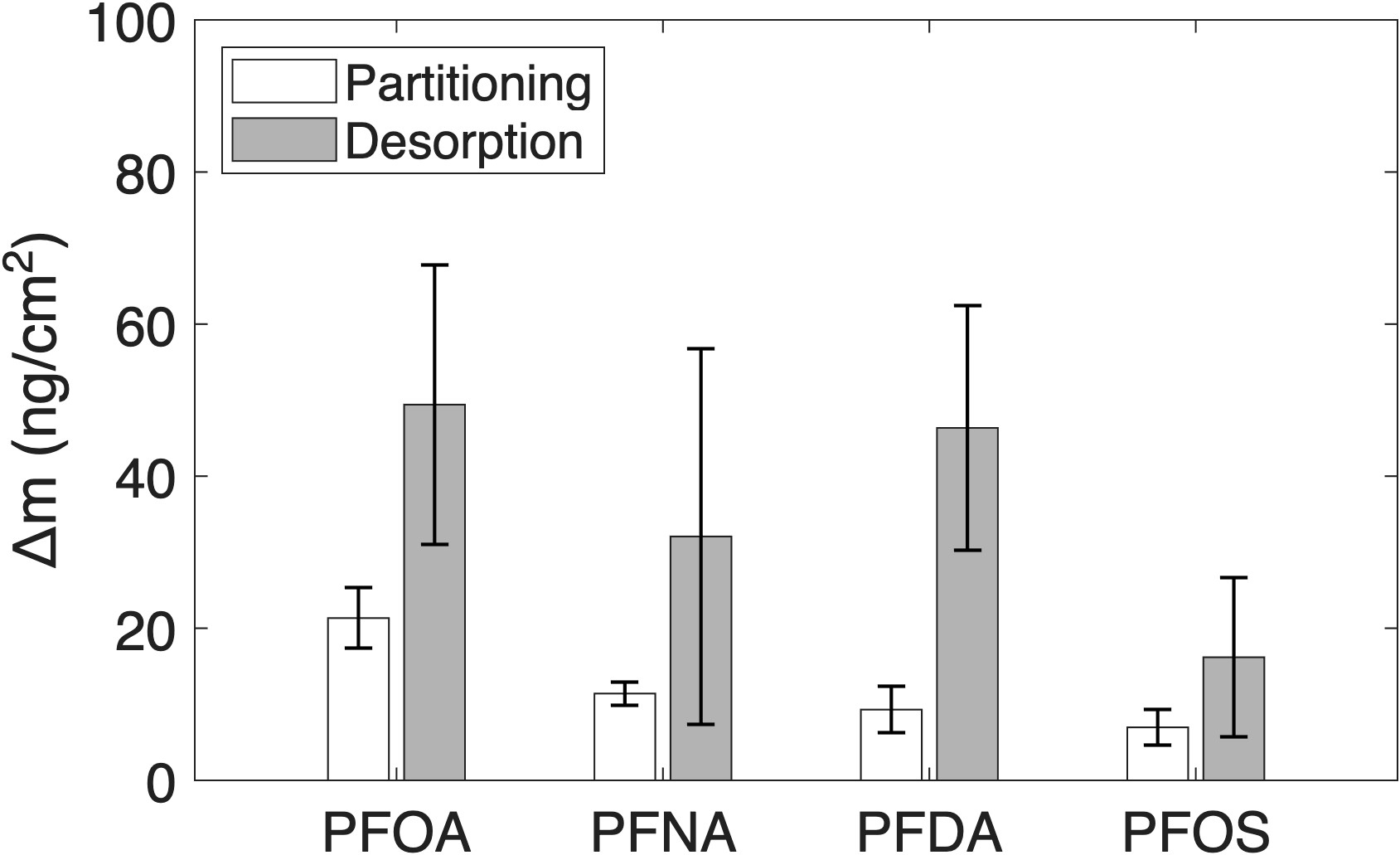}
	\caption*{Fig. S3:
		\textbf{Changes in surface mass density $\Delta m$ induced by PFAS partitioning and desorption.}}
	\label{fig:sauerbrey}
\end{figure}
$\newline$

Unlike the zeta potential response, the magnitude of the QCM-D frequency shift did not vary monotonically with PFAS chain length, and no clear trend was observed among the PFAS species. We attribute this lack of correlation to concurrent lipid loss during PFAS partitioning. Because QCM-D measures the total mass coupled to the silica sensor rather than detecting PFAS directly, it cannot distinguish the mass gained through PFAS partitioning from the mass lost through lipid removal. Although longer chain PFAS partition more strongly into the membrane, they also induce greater lipid loss, which can substantially offset the corresponding mass increase. Consistent with this interpretation, a previous neutron-reflectometry study found that exposure to 0.15~mM PFNA reduced the surface lipid density by approximately 30\% \cite{Nouhi2018}. Consequently, the net mass change measured by QCM-D does not directly quantify the extent of PFAS partitioning.

Although QCM-D alone cannot quantify PFAS partitioning or desorption, the measurements provide qualitative evidence that PFAS exposure causes substantial lipid loss from the supported lipid bilayer. Together with the zeta potential measurements, these results support the proposed mechanism by which PFAS induces lipid removal and consequently decreases the vesicle zeta potential.

\bibliography{reference}% Produces the bibliography via BibTeX.

\end{document}